\documentclass{ifacconf}

\usepackage{graphicx}      
\usepackage{natbib}        

\usepackage{amsmath,amsfonts,amssymb}
\usepackage{gensymb}
\usepackage{color}
\usepackage{subfloat}
\usepackage[american]{circuitikz}
\usepackage{todonotes}

\begin{document}
\begin{frontmatter}

\title{Li-ion Battery Fault Detection in Large Packs Using Force and Gas Sensors} 

\thanks[footnoteinfo]{This work has been accepted to IFAC for publication under a Creative Commons Licence CC-BY-NC-ND.}

\author[First]{Ting Cai} 
\author[First]{Peyman Mohtat} 
\author[First]{Anna G. Stefanopoulou}
\author[First]{Jason B. Siegel}

\address[First]{University of Michigan, 
   Ann Arbor, MI 48105, USA \\(e-mail: \{{tingcai, pmohtat, annastef, siegeljb\}}@umich.edu).}

\begin{abstract}                
Internal short circuits are a leading cause of battery thermal runaway, and hence a major safety issue for electric vehicles. An internal short circuit with low resistance is called a hard internal short, which causes a high internal current flow that leads to an extremely fast temperature rise, gas generation, cell swelling, and ultimately battery rupture and failure. Thus it is crucial to detect these faults immediately after they get triggered. In large battery packs with many cells in parallel, detecting an internal short circuit event using voltage is difficult due to suppression of the voltage signal from the faulty cell by the other healthy cells connected in parallel. In contrast, analyzing the gas composition in the pack enclosure can provide a robust single cell failure detection method. At elevated temperature, decomposition of the battery materials results in gas generation and cell swelling. The cell structure is designed to rupture at a critical gas pressure and vent the accumulated  $CO_2$ gas, in order to prevent explosive forces. In this paper, we extend our previous work by combining the models of cell thermal dynamics, swelling, and $CO_2$ gas generation. In particular, we developed a fast and high confidence level detection method of hard internal short circuit events for a battery pack by measuring cell expansion force and monitoring $CO_2$ concentrations in a pack enclosure.
\end{abstract}


\end{frontmatter}

\section{Introduction}
The continuous increase in Li-ion battery energy density is a necessary and important step to reduce the cost and range anxiety of electric vehicles (\cite{Feng2018}). The growth in energy density also increases the risk and severity of battery failures. Many battery accidents are triggered by overcharge, overheat, mechanical abuse (\cite{Feng2018}) or lithium plating that leads to battery internal short circuit (ISC). Joule heating caused by an internal short elevates the battery temperature. At elevated temperatures, battery side reactions become active and quickly produce heat that cannot be dissipated, which leads to battery thermal runaway (\cite{Spotnitz2003}), and ultimately fire and even an explosion (\cite{Abada2016}). Hence, the detection of a battery ISC event should be made early and accurately to execute emergency procedures and evacuate personnel.

Previous methods of detecting ISC are usually based on voltage or current measurements. Voltage based methods work well with a single cell, and experiments (see \cite{Feng2014, zhang2017fusing}) showed a significant battery voltage drop after the battery abuse tests. By using voltage measurement, \cite{xia2016fault} proposed a fault-tolerant method that can distinguish between cell failure and voltage sensor failure. The use of a correlation coefficient calculated for neighboring cells in series can result in a  model-free fault detection algorithm (\cite{xia2017correlation}). However, in large battery packs it can be difficult to identify the fault using only voltage measurements (\cite{Cai2020battery}). In electric vehicle battery packs, the cells are connected in parallel inside a module. For example, inside the Tesla Model S battery pack, there are 74 cells in parallel (\cite{bruen2016modelling}). For Tesla Model 3 battery pack, there are 46 cells in parallel. A large number of parallel-connected batteries will suppress the battery fault voltage signal. As the other healthy cells in parallel will continually supply nominal voltage, the pack voltage with one cell at fault will be similar to the voltage of healthy battery packs, making the fault detection using voltage alone challenging.



Recent studies have focused on ISC detection by integrating voltage, current, and surface temperature measurements (\cite{Feng2016, dey2017model,dey2019safer}). These fault detection methods work well with a soft internal short circuits, where the temperature gradient inside the cell is small. For hard internal short circuits, the battery internal temperature can be elevated in a few seconds, causing a large temperature gradient inside the cell. In \cite{Cai2018, cai2019modeling}, the authors divided the battery into three temperature sections and showed that at the early stage of ISC, the battery surface temperature rise is much slower than the voltage drop and the expansion force rise. In \cite{Lammer2017}, the authors showed that a large amount of $CO_2$ is released with the first venting during a thermal runaway event. Several papers have proposed detection methods based on sensing the vented gas during the thermal runaway process (\cite{liao2019survey, fernandes2018identification}). The gas detection method is advantageous when individual cell voltage and temperature measurements are not possible. For example, in a prior work from \cite{cai2019early}, for a battery storage drum, a gas detection method targeted at $CO_2$ concentrations shows a much faster detection speed than that from temperature monitoring at the drum surface.

For internal short circuit events that evolve without going into thermal runaway, cell surface temperature increase is limited (\cite{cai2019modeling}) and the fault is even more difficult to detect by conventional methods using voltage and temperature measurements. If left undetected, the cell might develop into a thermal runaway after continuous use (\cite{zhang2017fusing}). This type of event features a fast drop and quick recovery of the voltage and is named as the ``Fusing Phenomenon" in \cite{zhang2017fusing}. In this internal short, the high temperature in the ISC region will trigger battery side reactions, which produce a large amount of gas (\cite{cai2019modeling}). The generated gas leads to the swelling of the pouch cell that can be measured as a sudden increase in expansion force at the pack level. The generated gas can also be released in the event of a rupture that elevates the $CO_2$ concentration level inside the pack. The $CO_2$ level can be measured using a gas sensor inside the battery pack. 

The goal of this paper is to develop a high confidence short circuit detection method based on the measurement of cell expansion force and $CO_2$ level in the pack. To this end, we have developed an observer for the cell expansion in normal operating conditions to detect battery faults from force measurement. Furthermore, a $CO_2$ gas sensor is used to detect abnormal gas concentration spikes. The results indicate that in the absence of voltage measurements, the proposed algorithm can detect a hard short circuit quickly in a battery pack by monitoring force and gas levels.

\section{Battery Internal Short Circuit Model}
For electric vehicle packs, cells are connected with up to 74 cells in parallel, like in the Tesla Model S. Here, we consider a battery pack with 50 cells in parallel. 

\subsection{Terminal Voltage and Thermal Model}
\begin{figure}[ht]
\centering
\ctikzset{bipoles/length=0.9cm}
\begin{circuitikz}[american voltages]\draw
(0,1.5) to [V, v_=$V(SOC)$] (0,0);
\draw (0,1.5) to [R, l=$R_{cell}$] (0,3)
to [short] (2.2,3)
to [R] (2.2,1.5)
to [V] (2.2,0)
to [short] (0,0);

\draw (1.1,1.5) to [V] (1.1,0);
\draw (1.1,1.5) to [R] (1.1,3);

\draw (2.2,0) to [short] (2.9,0);
\draw (3.2,0) node[circ]{};
\draw (3.4,0) node[circ]{};
\draw (3.6,0) node[circ]{};

\draw
(3.9,0) to [short] (4.5,0);
\draw (4.5,1.5) to [V] (4.5,0);
\draw (4.5,1.5) to [R] (4.5,3);

\draw (2.2,3) to [short] (2.9,3);
\draw (3.2,3) node[circ]{};
\draw (3.4,3) node[circ]{};
\draw (3.6,3) node[circ]{};
\draw
(3.9,3) to [short] (4.5,3);

\draw (4.5,0) to [short] (5.5,0)
to  [R, l_=$R_{short}$, i<_=$I_{short}$] (5.5,3.0)
to [short] (4.5,3);
\end{circuitikz}
\caption{\small{Equivalent circuit model representing a battery pack with $n$ parallel connected cells and one cell with an internal short circuit.}}
\label{fig:ECM}
\end{figure}
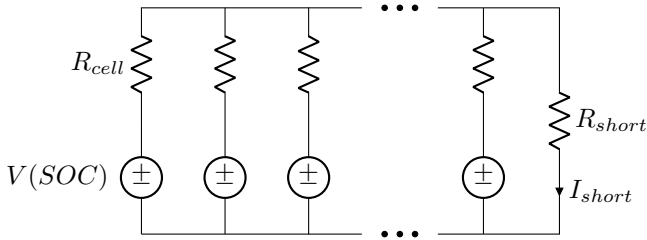

For a short circuit in a battery pack with $n$ cells in parallel, the equivalent circuit can be represented by Fig.~\ref{fig:ECM}. Here we assume the capacity for each cell is 4.5 Ah.

After triggering an internal short circuit, the major heat source is the ohmic heat from the internal short circuit current. The internal short current for the shorted cell and the terminal voltage can be written as
\begin{equation}
I_{short}=\frac{n \cdot V(SOC)}{R_{cell}+n \cdot R_{short}} \label{eq:IS}
\end{equation}
\begin{equation}
V_{T}=V(SOC)-I_{short}/n \cdot R_{cell} \label{eq:VT}
\end{equation}

where $V(SOC)$ is the open circuit voltage, which is a function of State of Charge ($SOC$),  $V_T$ is the terminal voltage, $I_{short}$ is the short circuit current, $R_{short}$ is the short circuit resistance, $R_{cell}$ is the cell impedance at 1 kHz, and $n$ is the number of parallel connected cells ($n=50$ in this case). By substituting Eq.~(\ref{eq:IS}) into Eq.~(\ref{eq:VT}), it is clear that for large $n$, the change in $V_T$ caused by an internal short is reduced.

The internal short circuit region is a small area inside the cell. The localized heating $Q_{ohmic}$  due to the ISC, causes a rapid local temperature increase, while temperature for the rest of the cell remains relatively constant during the early stages of thermal runaway (see \cite{cai2019modeling}). Above 120~$^\circ$C the Solid Electrolyte Interface (SEI) decomposition becomes active and starts to generate significant heat (\cite{Feng2018}). Here, we focus on modeling only the temperature of the ISC region $T_{ISC}$, and assume a constant $T_{cell}$ for the rest of the cell. The thermal model can be expressed as
\begin{equation}
C_p\frac{dT_{ISC}}{dt}=(Q_{SEI,ISC}+Q_{ohmic,ISC})\\+\frac{T_{cell}-T_{ISC}}{R_c}
\label{eq_ISCTemp}
\end{equation}
\begin{equation}
Q_{ohmic,ISC}=I_{short}^2R_{short}
\end{equation}
\begin{equation}
Q_{SEI,ISC}=-m_{an,ISC} \cdot h_{SEI} \cdot \frac{dx_{SEI,ISC}}{dt}
\label{eq_SEIQ}
\end{equation}
where $T_{ISC}$ and $T_{cell}$ represent the ISC region temperature and the cell temperature respectively. $R_c$ is the thermal resistance between the ISC region and the rest of the cell. $C_p$ is the thermal capacity of the ISC region. $Q_{SEI,ISC}$ is the reaction heat from SEI decomposition. $Q_{ohmic,ISC}$ is the ohmic heat in ISC region, $h_{SEI}$ is the reaction enthalpy of SEI decomposition, and $m_{an,ISC}$ is the mass of anode in the ISC region.

 
 The SEI decomposition reaction rate will increase exponentially with temperature (see \cite{Hatchard2001}), and can be expressed as
\begin{equation}
\frac{dx_{SEI,ISC}}{dt}=-A_{SEI} \cdot x_{SEI,ISC} \cdot \exp\left( -\frac{E_{SEI}}{k_b T_{ISC}} \right)
\label{eq_SEI}
\end{equation}

where $x_{SEI,ISC}$ is the fraction of Li in the SEI in the ISC region, representing the progress of SEI decomposition. $A_{SEI}$ is the frequency factor for SEI decomposition. $E_{SEI}$ is the activation energy for SEI decomposition, and $k_b$ is Boltzmann's constant.




 \subsection{Gas Generation Model}
The SEI decomposition reaction generates gas that can lead to severe cell swelling and venting of gas. The total amount of $CO_2$ generated can be expressed as a function of SEI decomposition progress $x_{SEI,ISC}$
\begin{equation}
n_{CO_2}=\frac{m_{an,ISC}(x_{SEI,0}-x_{SEI,ISC})}{2M_{C_6}}
\label{gas_evolution}
\end{equation}

where $n_{CO_2}$ is the amount of $CO_2$ in mole, $x_{SEI,0}$ is the initial $x_{SEI}$ before side reactions become active, and $M_{C_6}$ is the mass per mole ($g/mol$) for $C_6$. Model parameter values can be found in \cite{cai2019modeling}.






\subsection{Expansion Force Measurements in Battery Packs}

\begin{figure}[ht]
\centering
\includegraphics[width=1.0\columnwidth,trim = 0.2in 0in 0in 0in,clip=true]{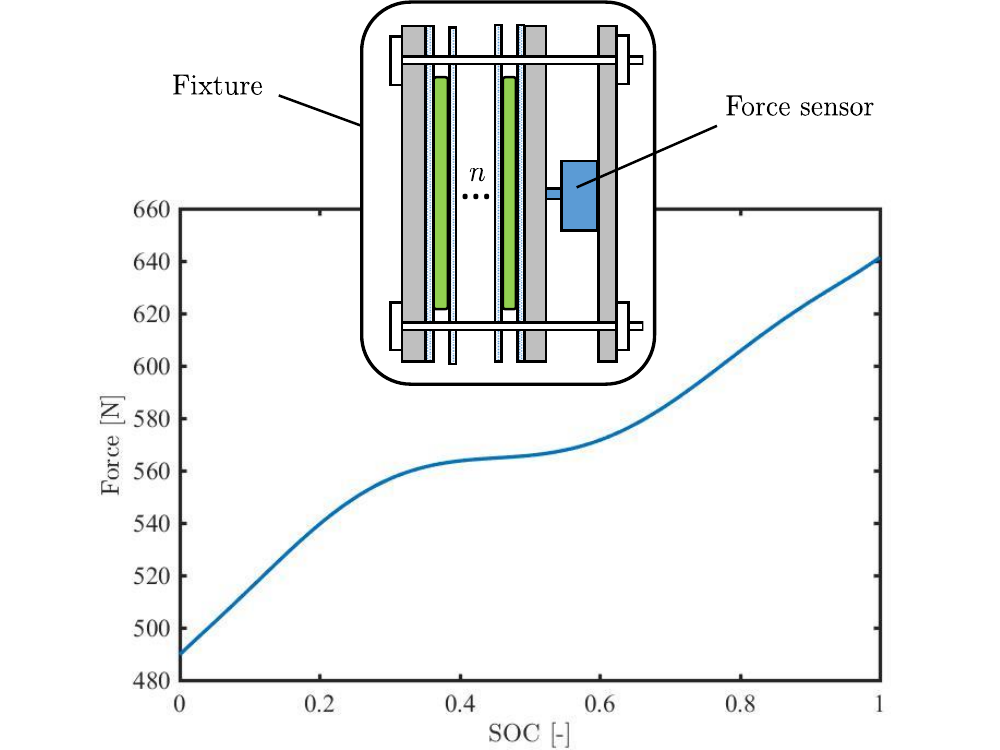}
  \caption{\label{Force_Setup} \small{Expansion force as a function of SOC for a NMC prismatic pouch 4.5 Ah cell at 25\degree C, with the schematic of the force sensor placement in a pack.}}
\end{figure}

In a battery pack, the cell expansion due to changes in SOC, internal gas pressure, and cell temperature during normal operation and fault conditions should be considered in the model. For automotive battery packs, the cells are typically constrained to a fixed volume as shown in the inset of Fig.~\ref{Force_Setup}. Therefore swelling of the cell would result in an increase in the cell volume, which would tend to exert a force that is balanced by the pack end plates. This change in force can be measured for multiple cells, which are mechanically connected in series, and is shown in Fig.~\ref{Force_Setup}. This figure illustrates the change in the force during nominal operation for a discharge cycle.



\subsection{Expansion Force Model}

At normal operating conditions, the cell expansion force can be expressed as a function of temperature and State of Charge (SOC). For a single cell, the change of expansion force is around 156 N or 30\% of the total force from a fully discharged to a fully charged state. The peak force due to an internal short circuit event can exceed the sensor capacity (1780 N) and hit the sensor saturation limit of 3560 N (see \cite{cai2019modeling}), which is greater than 10x of the normal expansion force change. Here, we model the expansion force as a separable function of temperature and SOC (see \cite{mohan2014phenomenological}). The expansion force can then be expressed as
\begin{equation}
F=f_1(T)+f_2(SOC)+F_0+F_{gas}
\label{Force}
\end{equation}
where the $F_{gas}$ term is the fault expansion force due to gas generation and $F_0$ is the preload force. For the SOC dependency of the expansion force, the measurement experiment uses the same setting as \cite{poloni2018integration}. The cell chemistry is nickel manganese cobalt oxide (NMC), with prismatic structure. Here, an eighth order polynomial fit for the expansion force as a function of SOC is used as shown in Fig \ref{Force_Setup}.

For the temperature dependence, we assume the expansion force grows linearly with the temperature as 
\begin{equation}
f_1(T)=\alpha (T-T_0)
\end{equation}
where $T_0$ is the initial temperature, and $\alpha$ is the thermal expansion rate.The thermal expansion coefficient $\alpha$ varies with different fixtures and batteries. Here, we calculate it based on the experimental data for the cell during the heating phase. The $\alpha$ in this study is $2.06~N/ \degree C$.



As a result of gas generation, the cell swells and increases the measured expansion force once the gas pressure inside the cell overcomes the preload force. The increased force due to generated gas is modeled using the ideal gas law to convert the number of moles of gas to pressure. We can express the pressure as
\begin{equation}
P=\frac{n_{CO_2}RT_{cell}}{\Delta V}=\frac{F_{gas}}{A_{cell}}
\label{eq:Pressure}
\end{equation}

where $\Delta V$ is the change of cell volume occupied by the gas. We assume $\Delta V=A_{cell} \Delta x$,  and the deflection of the cell casing is balanced by the increased force from the fixture and compression of adjacent cells $F_{gas}=K_{eq} \Delta x $. An equivalent spring constant of the battery pack $K_{eq}$ is used which captures the effects of all other cells in the pack. Since $K_{eq}$ is the result of a series connection of mechanical springs, it is expected that the value would decrease as the number of cells in the pack increases or as the pre-load decreases. We plug these relationships into Eq.~(\ref{eq:Pressure}) to solve for the change in cell thickness $\Delta x$ in the direction of the applied force. 

Hence the fault expansion force $F_{gas}$ can be expressed as
\begin{equation}
F_{gas}= \sqrt{K_{eq} \cdot n_{CO_2} \cdot RT_{cell}}
\label{Force_evolution}
\end{equation}
The expansion grows rapidly following the SEI decomposition. Hard cased cylindrical and prismatic battery cells are designed with a venting structure that will reliably fail open once the cell exceeds an internal pressure of 3448 kPa to prevent explosive forces due to gas buildup (see \cite{coman2016lumped}). After the cell ruptures, the fault force drops to zero as a consequence of the release of gas.


\subsection{Gas Concentration in the Pack}

After the fault event, the vented gas transport process is fast, and \cite{said2019comprehensive} indicated a transport time of 3 seconds for $CO_2$ sensors. To model the gas sensor response which is located at the pack vent-gas duct outlet, a 1D mass transport equation is incorporated. The model assumes diffusion and convection processes. Here, we assume the vent-gas velocity prescribes the airflow velocity. The length of the total battery pack is assumed $0.5~m$. The $CO_2$ is assumed to be generated at the boundary location ($x=0$) for the duration of the gas venting event. The mass transport equation is as the following
\begin{equation}
\frac{dc}{dt}=-\frac{\partial}{\partial x}\left(-D\frac{\partial c}{\partial x} + cv\right)+r
\label{mass transport}
\end{equation}
where $c$ is the concentration of $CO_2$. $D=14.2~mm^2/s$ is the diffusion coefficient of $CO_2$ in the air. $v(x,t)$ is the vent-gas velocity distribution as a function of location ($x$) and time ($t$), and follows the equation below
\begin{equation}
v(x,t)= \begin{cases}
    v_0, & \text{if $x> v_0(t-t_0)$ \& $x<v_0t$}\\
    0, & \text{otherwise}.
  \end{cases}
\label{Velo_distrib}
\end{equation}
where $v_0$ is the initial vent velocity, and can be derived using the amount of gas and the duration of the gas venting
\begin{equation}
v_0 = \frac{n_{CO_2}RT_{gas}}{PA_{Rupture}t_0}=0.12~m/s
\end{equation}
where $A_{Rupture}$ is assumed to be the area of the rupture, $P$ the atmospheric pressure, $R$ the gas constant, $T_{gas}$ the average gas temperature, and $t_0$ the duration of the gas venting ($t_0= 1.5~s$ from simulation using model of \cite{cai2019modeling}). The source term $r$ for the $CO_2$ generation follows the equation below
\begin{equation}
r = \frac{n_{CO_2}}{A_{Rupture}h_{ch}t_0}
\label{gas gen}
\end{equation}
where $h_{ch}$ is the pack vent-gas channel height. The initial concentration is set to 400 ppm. Furthermore there is a Dirichlet boundary condition at outlet, which corresponds to the atmosphere $CO_2$ concentration, $c(x_{outlet},t)=400$ ppm. 





\section{Fault Detection Methodology for Packs}

Prior research from \cite{Cai2020battery} showed the slow response of using surface temperature to detect an ISC event. This is primarily due to the poor observability of using surface temperature measurements to estimate the core temperature state where the ISC happens. As discussed in the previous section, with a large number of parallel-connected cells in a battery pack, using voltage measurements is difficult to identify a battery fault. Considering the fast response of the expansion force signal and the gas concentrations signal, we propose an ISC detection methodology for battery packs based on expansion force measurements and gas sensing.

\subsection{Fault Detection Algorithm Using Expansion Force}
From the above discussions, an expansion force model is built during normal operating conditions. Based on this model, we then build an observer for the expansion force
\begin{equation}
\hat{F}=f_1(T)+f_2(\hat{SOC})+F_0+\hat{\Theta}
\label{Force_estimate}
\end{equation}
\begin{equation}
\dot{\hat{\Theta}}=L(\bar{F}-\hat{F})
\label{bias_L}
\end{equation}
where $\bar{F}$ is the measured force, and $\hat{\Theta}$ is the estimated fault force signal. $\hat{\Theta}$ can be derived from force measurement and the estimated expansion force.


For SOC estimation, Coulomb Counting and Open Circuit Voltage inversion are the two main methods. The SOC estimation error will influence the $\hat{\Theta}$. For a single cell, a $5\%$ SOC estimation error can contribute about $8.9~N$ error for $\hat{\Theta}$. The estimation error can increase due to sensor noise, sensor drift and model mismatch due to cell aging (see \cite{mohtat2019towards}). Closed-loop SOC estimation is needed with Kalman Filter (\cite{plett2004extended}) to balance between process error and sensor noise and achieve less SOC estimation error. In this study, we will use the Coulomb Counting for SOC estimation for simplicity. 

At normal operating conditions, the measurement for expansion force should match the model, and ideally $\Theta$ should be zero. However, $\Theta$ will not necessarily be zero due to modeling error and sensor noise. In a short circuit case, after the expansion force surges in a few seconds, the estimated fault force signal $\Theta$ will increase rapidly due to the error correction term. To summarize the two cases:

\textbf{During Normal Conditions}
\begin{equation*}
\bar{F}=f_1(T)+f_2(SOC)+F_0
\end{equation*}
\begin{equation*}
\hat{F}=f_1(T)+f_2(\hat{SOC})+F_0+\hat{\Theta}
\end{equation*}
\begin{equation*}
\hat{\Theta} \rightarrow 0
\end{equation*}

\textbf{At Fault Conditions}
\begin{equation*}
\bar{F}=f_1(T)+f_2(SOC)+F_0+F_{gas}
\end{equation*}
\begin{equation*}
\hat{F}=f_1(T)+f_2(\hat{SOC})+F_0+\hat{\Theta}
\end{equation*}
\begin{equation*}
\hat{\Theta} \rightarrow F_{gas}
\end{equation*}
where the $F_{gas}$ term represents the increased expansion force as a result of abnormal cell swelling. During normal operating conditions, the $\hat{\Theta}$ converges to zero. In a short circuit event, the detection quantity $\hat{\Theta}$ converges to $F_{gas}$. Thus, the detection algorithm can be written as
\begin{equation}
\left|\hat{\Theta}\right|   > \epsilon_F,~~~~~ \text{ISC Alert}.
\label{detection_criteria}
\end{equation}
where $\epsilon_F$ is the predefined threshold for the expansion force. A small threshold may lead to improved detectability of the fault but could cause false alarms.

\subsection{Higher Confidence Level Detection with Gas Sensor}
ISC detection methods based on a single sensor may suffer from sensor error, which will lead to undesired false alarms. Here, for higher confidence level, we use expansion force measurement and gas concentration measurement for ISC detection. If only one signal indicates a fault, then this might be a sensor error. If both signals indicate a fault then sensor failure can be ruled out and the cause is most likely an ISC event. With multiple detection algorithms, we can decrease the number of false alarms.


For the $CO_2$ gas concentration, we define the fault gas concentration value as
\begin{equation}
G_{fault}=G_{normal}-\bar{G}
\end{equation}
where $\bar{G}$ is the measured $CO_2$ concentrations in ppm, and $G_{normal}$ is the normal $CO_2$ gas concentrations in atmosphere, which is set as 400 ppm in this study.

If the fault gas concentration value $G_{fault}$ exceeds the pre-defined value, $\epsilon_G$, then the gas detection system will trigger an alarm. In the system with both force and gas sensors, only after receiving alarms from both detection systems in a short time frame, an ISC event is believed to have occurred.

\begin{table}[ht]
 \caption{Detection Logic}
 \begin{center}
 \begin{tabular}{l l l }
  \hline
 $CO_2$ Concentrations &  Force & Decision\\
 \hline
 $G_{fault}>\epsilon_G$ &  $\left|\hat{\Theta}\right|>\epsilon_F$ & ISC Alert\\

$G_{fault}>\epsilon_G$ &  $\left|\hat{\Theta}\right|<\epsilon_F$ & Warning\\

$G_{fault}<\epsilon_G$ &  $\left|\hat{\Theta}\right|>\epsilon_F$ & Warning\\

$G_{fault}<\epsilon_G$ &  $\left|\hat{\Theta}\right|<\epsilon_F$ & Normal\\
  \hline
 \end{tabular}
 \end{center}
 \end{table}
\section{Simulation Result}
For this study, we consider a battery pack with 50 parallel connected 4.5 Ah NMC pouch cells. The model parameters are adopted from \cite{cai2019modeling}. 
\subsection{Simulation Settings}
Zero mean white Gaussian noise ($N(0,\sigma^2)$) is added to the measurement to emulate a real system. The covariance of the noise for the voltage measurement is $\sigma_V=5~mV $. For the current measurement, $\sigma_I= 5~mA$ (see \cite{dey2019safer}). For the temperature measurement, $\sigma_T=0.5~ \degree C$ (Omega K-type thermal couple). For the force measurement, $\sigma_F=8.9~N$ (Omega). For the gas measurement, $\sigma_G=30~ ppm$ (Amphenol).

The Urban Dynamometer Driving Schedule (UDDS) is used for the current profile. Before triggering the fault, the pack operates under the UDDS cycle without a fault. Then, an internal short circuit is triggered at $t=10~s$, which shuts down the cell and disconnects the ISC current path 0.4 seconds later. The pack continues to operate under the UDDS cycle after the fault.

In the following simulation, we will use the Coulomb Counting method to estimate SOC, which is purely based on the current measurement. For the detection threshold, considering the sensor measurement error, the gas detection threshold is set to $\epsilon_G=2000~ppm$, and the force detection threshold is set to $\epsilon_F=100~N$. 






\subsection{Simulation at Fault Conditions}
In this simulation for the battery pack, a hard internal short circuit is triggered in a cell. The cell triggers ISC at $t=10~s$ with a short circuit resistance $R_{short}=25~m\Omega$. The fast short circuit process is stopped after the ISC current path is burnt down (see \cite{zhang2017fusing}). The voltage quickly returns to normal and there is no significant surface temperature increase for such an event. The cell swells and ruptures after 1.5 seconds of the ISC initialization. Although this fault will not directly lead to thermal runaway at this time, a second-time ISC might occur soon, so the event needs to be identified early to safely handle the battery pack with the faulty cell.

The simulated hard short circuit event is shown in Fig.~\ref{simu_fault} for the current and voltage, and Fig.~\ref{simu_fault_Force} for the force and gas concentrations profile. The first 10 seconds simulation is free of fault, and both detection quantities are below the threshold. The short circuit fault triggers at $t=10~s$. Note that from Fig.~\ref{simu_fault}, it is difficult to identify the fault with voltage measurements for a battery pack.



\begin{subfigures}
\begin{figure}[ht]
\centering
\includegraphics[width=0.92\columnwidth,trim = 0.in 1.3in 0.in 1.3in,clip=true]{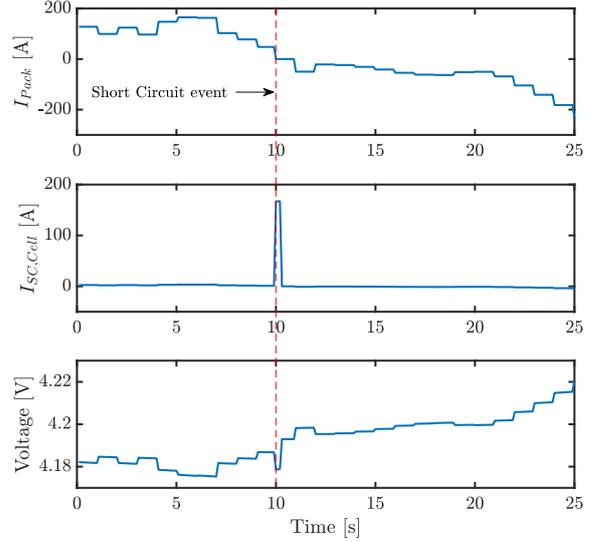}
  \caption{\label{simu_fault} \small{Pack current, short circuit cell current and voltage profile under a fault condition, with a hard short circuit triggered at $t=10s$. Note that no significant change for pack current and voltage is observed.}}
\end{figure}

\begin{figure}[ht]
\centering
\includegraphics[width=0.92\columnwidth,trim = 0.in 2.5in 0.3in 2.5in,clip=true]    {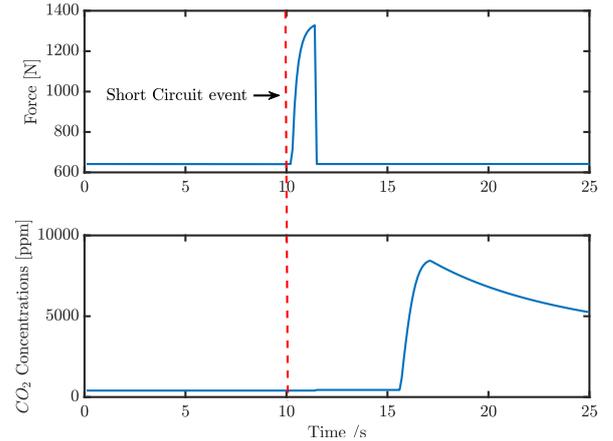}
  \caption{\label{simu_fault_Force} \small{Expansion force and gas concentration at the pack outlet after a short and cell rupture, with an internal short circuit triggered at $t=10s$.}}
\end{figure}

\end{subfigures}

\begin{figure}[ht]
\centering
\includegraphics[width=0.9\columnwidth,trim = 0.in 2.5in 0.3in 2.5in,clip=true]{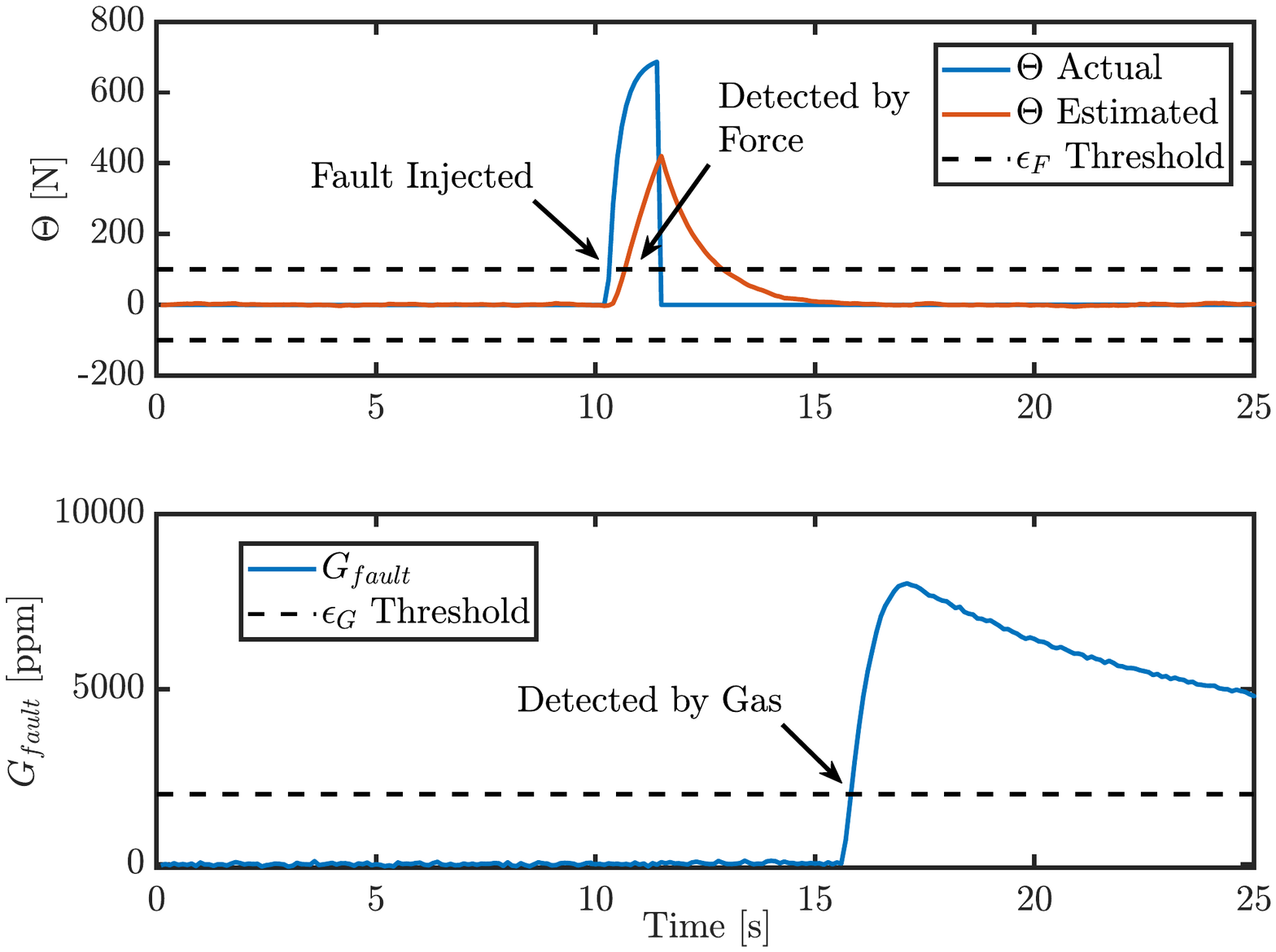}
  \caption{\label{Theta_fault} \small{At fault conditions, force detection $\hat{\Theta}$ identifies a fault at $t=10.7s$, and gas detection $G_{fault}$ confirms the fault at $t=15.9s$.}}
\end{figure}

The estimated gas fault term $G_{fault}$, estimated force fault term $\hat{\Theta}$ and the actual $\Theta$ after a short circuit triggered are shown in Fig.~\ref{Theta_fault}. At $t=10.7~s$, the force detection algorithm identifies the fault, and the gas sensor confirms the event at $t=15.9~s$. Even though the confirmation of an ISC event requires threshold crossing from both force and gas signals, it still achieves fast detection for a hard internal short event.


\section{Conclusion}
In this paper, we propose a battery internal short circuit detection method based on battery expansion force measurement and gas sensing. The study primarily focuses on a specific type of ISC event that features a fast voltage drop and recovery, and no significant change in surface temperature. This event is difficult to be identified using voltage and surface temperature measurements. 

The simulation shows a fast response for an ISC event based on the proposed method. The threshold values for both gas and force detection quantities are chosen manually, and this requires further study for optimal threshold values. Also, the presented simulation doesn't take into account sensor drift or model mismatch for aged cells, which can cause a bias in the estimation and can lead to error in the detection quantities. Further study is required to properly handle the issues of bias and drift.



\begin{ack}
{This work is supported by the National Science Foundation under Grant No. 1762247. The authors wish to acknowledge Miriam Figueroa for assistance with UDDS current profile, and Vivian Tran for improving the write-up of the paper. The authors wish to thank Brian Engle from Amphenol company for providing the gas sensor.}
\end{ack}

\bibliography{root}             
                                                   







\end{document}